\documentstyle[aps,preprint]{revtex}
\begin{document}
\draft{}
\title{  Analysis of preliminary data on 
$e^+e^-\to\phi\to\gamma f_0(980)\to\gamma\pi^0\pi^0$ reaction.}

\author{N.N. Achasov
\thanks{ E-mail: achasov@math.nsc.ru} and  V.V. Gubin
\thanks{ E-mail: gubin@math.nsc.ru }}
\address{Laboratory of Theoretical Physics\\
S.L. Sobolev Institute for Mathematics\\
630090 Novosibisk 90,\  Russia}
\date{\today}
\maketitle
\begin{abstract}

We perform the analysis of the preliminary data on
$e^+e^-\to\phi\to\gamma f_0(980)\to\gamma\pi^0\pi^0$ reaction 
simultaneously with the data on $\pi\pi$ scattering and   reactions
$J/\psi\to\phi\pi^+\pi^-$ and $K^-p\to\pi^+\pi^-(\Lambda,\Sigma)$.
It is found that the $f_0(980)$ meson
mass  $m_{f_0}=950$ MeV and  
$B(\phi\to\gamma f_0\to\gamma\pi^0\pi^0)\simeq1\cdot10^{-4}$.

\end{abstract}


 The peculiar properties of the $a_0(980)$ and $f_0(980)$ mesons has 
 been the center of attention for years. It is well known that
 the standard quark model does not account for all properties of  
 the $a_0(980)$ and $f_0(980)$ mesons, see, for example \cite{achasov-84}.
 In time, all their challenging properties could be understood in the 
 framework of the four-quark ($q^2\bar q^2$) MIT-bag model \cite {jaffe-77},
 see \cite{achasov-84}. Besides, along with the  $q^2\bar q^2$ nature
 the possibilities of them being the $K\bar K$ molecules \cite{weinstein} 
 and traditional $q\bar q$ states \cite{tornqvist} ( the $s\bar s$ model
for $f_0$ meson) are discussed in the  literature.
 As it was established in papers
  \cite{achasov-89,isgur-93,molecule,neutral} the radiative decays of
the  $\phi$ meson ($\phi\rightarrow\gamma f_0\rightarrow\gamma\pi\pi$ and
$\phi\rightarrow\gamma a_0\rightarrow\gamma\eta\pi$) could be a good 
guideline in distinguishing of the $f_0$ and $a_0$ meson models.
In this connection the active experimental investigation of this decays has
been carried out with Spherical Neutral Detector (SND) and
Cryogenic Magnetic Detector-2 (CMD-2) at $e^+e^-$-collider
VEPP-2M in Novosibirsk and has been planed at $\phi$-factory  DA$\Phi$NE in
Frascati and at CEBAF. The preliminary data on this decays already 
has been obtained with SND,
 see \cite{snd}.
It follows  from experiment that  $B(\phi\to\gamma\pi^0\pi^0)=
(1.1\pm0.2)\cdot10^{-4}$ and
 $B(\phi\to\gamma\pi\eta)=(1.3\pm0.5)\cdot10^{-4}$ for the photon energy
 $\omega<200$ MeV that points to the $(q^2\bar q^2)$ nature of the $a_0$ and
 $f_0$ mesons \cite{achasov-89,neutral}. The reasonable statistics 
lets  to draw the $\pi\pi$ mass spectra of the reaction
 $e^+e^-\to\phi\to\gamma\pi^0\pi^0$.  The analysis of experimental data
   on this reaction \cite{snd} was carried out with help of the following
   formula (see $g(m)$ and details in \cite{achasov-89,neutral})
\begin{eqnarray}
\label{signalfora0}
\frac{d\sigma_{\phi}}{dm}\sim\frac{\omega|g(m)|^2}
{\left|D_{f_0}(m)\right|^2}\sqrt{1-\frac{4m_{\pi}^2}{m^2}},
\end{eqnarray}
where $m$ is the mass of $\pi\pi$ system, $s$ is the square of the 
total energy of $e^+e^-$ beams,
 the energy of photons  $\omega=(s-m^2)/2\sqrt{s}$. 
$1/D_{f_0}(m)$ is the propagator of the $f_0$ meson,
\begin{equation}
\label{propagator}
D_{f_0}(m)=m_{f_0}^2-m^2+\sum_{ab}[Re P_{f_0}^{ab}(m_{f_0}^2)-
P_{f_0}^{ab}(m)].
\end{equation}
The sum $\sum_{ab}[Re P_{f_0}^{ab}(m_R^2)-P_{f_0}^{ab}(m)]$ takes into 
account the finite width corrections of resonance connected with channels of
$\pi\pi$,  $K\bar K$, $\eta\eta$, $\eta\eta'$ and etc.
 For the pseudoscalar  $ab$ mesons and $m_a\geq m_b,\ m^2>m_+^2$ one has
\begin{equation}
\label{polarisator}
P^{ab}_{f_0}(m)=\frac{g_{f_0ab}^2}{16\pi}\left[\frac{m_+m_-}{\pi m^2}\ln
\frac{m_b}{m_a}+\rho_{ab}\left(i+\frac{1}{\pi}\ln\frac{\sqrt{m^2-m_-^2}-
\sqrt{m^2-m_+^2}}{\sqrt{m^2-m_-^2}+\sqrt{m^2-m_+^2}}\right)\right].
\end{equation}
In other regions of $m$ one can obtain the $P_{f_0}^{ab}(m)$ by
analytical continuation.

Fitting of experimental data on   $e^+e^-\to\phi\to\gamma\pi^0\pi^0$ reaction
with described formula has given the following parameters of the $f_0$ 
meson \cite{snd} 
\begin{eqnarray}
\label{ivan}
&&m_{f_0}=950\pm8\ MeV,\quad g^2_{f_0K^+K^-}/4\pi=(2.3\pm0.5)\ GeV^2\nonumber\\
&&g^2_{f_0\pi^+\pi^-}/4\pi=(0.4\pm0.1)\ GeV^2
\end{eqnarray}
that corresponds to the  $q^2\bar q^2$ model \cite{achasov-89,neutral}.
The effective width of the $f_0$ meson for  this parameters is
 $\Gamma_{eff}\simeq60$ MeV ( the definition of the effective width see in
  \cite{neutral}).
Note that for this fitting $\chi^2=4.6$ at 7 degrees of freedom.

 The obtained from fitting  $f_0$ meson mass is relatively low. 
This raised the active discussion on HADRON-97 conference.
 It was pointed there that the mass $m_{f_0}=950$ MeV is noticebly lower 
than that presented by Particle Data Group \cite{pdg}. 
But, one should keep in mind that the mass in Eq.(\ref{ivan})
 is treated, in contrast to Particle Data Group, a la' field
theory, i.e. as the position of an inverse propagator real part zero
 ( as well as, for example, the $Z_0$ boson mass) and, correspondingly, 
this mass is the physical one of 
resonance.  The propagator pole for fitting (\ref{ivan}) is situated at
 $m_{f_0}^p=0.988-i0.08$ GeV that is in good agreement with
Particle Data Group \cite{pdg}.
 
 But there is another problem. The single resonance 
with mass $m_{f_0}=950$ MeV, as it was presented in fitting (\ref{ivan}),
 cannot describe the  $\pi\pi$ scattering data even one takes into
 account the elastic background with additional phase 
 $\theta\simeq80-90^{\circ}$. The $\pi\pi$ scattering data in the interval
 $0.7<m<1.1$ GeV can be described by only the "heavy" $f_0$ meson with mass
  $m_{f_0}=980$ MeV. To describe the $\pi\pi$ data with the "light" 
$f_0$ meson  one needs an extra $\sigma$ resonance.
 
 These things considered, we present the results of the mass spectra
 analysis of the reaction $e^+e^-\to\phi\to\gamma\pi\pi$
 included the simultaneous two-channels description of the spectra of the
 $e^+e^-\to\phi\to\gamma(f_0+\sigma)\to\gamma\pi\pi$ reaction and 
the data on the $\pi\pi$ scattering.
 Besides, we analyze the way of consistency of the $f_0$ meson parameters
obtained from our fitting  with the other available experimental data,
 i.e. the data on the $f_0$ meson production in the $J/\psi$ decays
 and the data on the $K^-p\to\pi^+\pi^-(\Lambda,\Sigma)$ reaction. 

 Our analysis rests on the previous papers, see \cite{achasov-84,neutral}.
We describe the $\pi\pi$ scattering data by two-channels model in which
the broad  ($\Gamma_{\sigma}\simeq300$ MeV) and relativly heavy 
($m_{\sigma}\simeq1.5$ GeV) resonance is considred besides the $f_0$ meson. 
To fit the $\pi\pi$ scattering data we write the $s$-wave amplitude of the
 $\pi\pi\to\pi\pi$ reaction with $I=0$ as the sum of the inelastic resonance
amplitude $T^{res}_{\pi\pi}$, in which the contribution of the $f_0$ and
 $\sigma$ mesons are taken into account,  and the amplitude of the 
elastic background \cite{achasov-84,neutral}
\begin{equation}
\label{eqpipi}
T(\pi\pi\to\pi\pi)=\frac{\eta^0_0e^{2i\delta^0_0}-1}{2i\rho_{\pi\pi}}=
\frac{e^{2i\delta_B}-1}{2i\rho_{\pi\pi}}+e^{2i\delta_B}T^{res}_{\pi\pi},
\end{equation}
where
\begin{equation}
\label{amplitudapipi}
T^{res}_{\pi\pi}=\sum_{RR'}\frac{g_{R\pi\pi}g_{R'\pi\pi}}{16\pi}
G^{-1}_{RR'}(m)
\end{equation}
The elastic background phase $\delta_B$ is taken in the form
 $\delta_B=\theta\rho_{\pi\pi}$, where $\theta\simeq60^{\circ}$.
 The matrix of the inverse propagator $G_{RR'}$ has the form
\begin{displaymath}
{G_{RR'}(m)}=
\left( \begin{array}{cc}
D_{f_0}(m)&-\Pi_{f_0\sigma}(m)\\
-\Pi_{\sigma f_0}(m)&D_{\sigma}(m)
\end{array} \right)
\end{displaymath}
 Nondiagonal elements of the matrix $G_{RR'}(m)$ are the transitions 
caused by the resonance mixing due to the final state interaction which
occured in the same decay channels  $R\to (ab)\to R'$. We write them down
in the following manner \cite{achasov-84,neutral}
\begin{equation}
\Pi_{RR'}(m)=\sum_{ab}\frac{g_{R'ab}}{g_{Rab}}P_R^{ab}(m)+C,
\end{equation}
where the constant $C$ takes into account effectively the contribution
of $VV,\ 40^-$ and other intermidiate states and incorporate
the subtraction constant for $R\to(0^-0^-)\to R'$ tarnsitions.
 In the four-quark model we treat this constant as a free parameter.
In this paper we take into account only the  $\pi\pi$ and $K\bar K$
intermidiate states in the matrix of inverse propagator. The consideration
of the other states does not change our results actually.

Making simultaneous fit of the $\pi\pi$ scattering data and the mass spectra
in the  $e^+e^-\to\phi\to\gamma(f_0+\sigma)\to\gamma\pi\pi$ reaction,
see Fig.1, we find that the best $\chi^2=6.2$
 for the mass spectra of the reaction
 $e^+e^-\to\phi\to\gamma\pi\pi$ is achieved at the following 
 $f_0$ meson parameters
 \begin{eqnarray}
\label{my}
&&m_{f_0}=950\ MeV,\quad g^2_{f_0K^+K^-}/4\pi=2.25\ GeV^2,\nonumber\\
&& R=g^2_{f_0K^+K^-}/g^2_{f_0\pi^+\pi^-}=3.5.
\end{eqnarray}
The effective width of the $f_0$ meson is $\Gamma_{eff}\simeq80$ MeV. 
The branching ratio in the region  $\omega<200$ MeV is
 $B(\phi\to\gamma(f_0+\sigma)\to\gamma\pi^0\pi^0)=1.03$.
The propagator pole for parameters in Eq.(\ref{my}) is situated at
$m_{f_0}^p=0.998-i0.14$ GeV.
Note that we cannot describe simultaneously the data  at masses
 $m_{f_0}>960$ MeV. For $m_{f_0}=960$ MeV ( in this case $R=4.0$ and the 
 other parameters are the same) we have  $\chi^2=8.6$. The total number of
 parameters in our model is 7 but we fit the three characters: the phase
 and inelasticity of the $\pi\pi$ scattering and the mass spectra of the 
 reaction $e^+e^-\to\phi\to\gamma\pi\pi$ (the total number of points is 110).
 The quantity  $\chi^2$ is presented for the $\pi\pi$ spectra in the
 $\phi\to\gamma\pi\pi$ decay only. So, our analysis gives actually the
 same parameters as in  Eq.(\ref{ivan}). 

 Besides, to clarify the question whether obtained in our fitting parameters
of the $f_0$ meson are consistent with the other experiment we analyze the mass 
spectra in the $f_0$ meson region in the 
$J/\psi\to\phi\pi^+\pi^-$ \cite{falvard} decay. In this decay
the mass spectra is determined by the following expression \cite{neutral}
\begin{equation}
\label{spectrjpsipipi}
\frac{dN_{\pi\pi}}{dm}=C\frac{m^2\Gamma_{f_0\pi\pi}(m)}{|D_{f_0}(m)|^2}
\cdot\left|\frac{ D_{\sigma}(m)+(1+\xi)\Pi_{f_0\sigma}(m)+
\xi(g_{\sigma\pi\pi}/g_{f_0\pi\pi})D_{f_0}(m)}
{D_{\sigma}(m)-\Pi_{f_0\sigma}^2/D_{f_0}(m)}\right|^2,
\end{equation}
where it has only two unknown parameters: $\xi$ is the relative weight of
the $\sigma$ meson direct production and $C$ is the overall coefficient. 
Fitting this parameters we find that the mass spectra can be well described
 with parameters obtained in the previous fitting, see Fig.2. For the set 
of parameters corresponding to $m_{f_0}=950$ MeV we have  $\xi=0.1$,
 $C=9.0$ and $\chi^2=19.7$  ( points number 23 ). For the set of parameters
corresponding to $m_{f_0} =960$ MeV we have $\chi^2=27.3$.

 So, the data on the $J/\psi\to\phi\pi^+\pi^-$ decay support the value
$m_{f_0}=950$ MeV.  Note that the data on the $J/\psi\to\phi\pi^+\pi^-$
decay cannot be described well by the single $f_0$ meson with the mass
$m_{f_0}=980$ MeV but these data are described by the single $f_0$ resonance
with mass $m_{f_0}=950$ MeV.

 In a similar manner we analyze the $\pi\pi$ spectra of the  
$K^-p\to\pi^+\pi^-(\Lambda\Sigma)$ reaction \cite{brandenburg}, see Fig.3.
The mass spectra in this reaction is determinated also by
 Eq. (\ref{spectrjpsipipi}). The data on $K^-p\to\pi^+\pi^-(\Lambda\Sigma)$ 
reaction are poorer than the data on the $J/\psi\to\phi\pi^+\pi^-$ decay
and consequently are not sensitive to the  $f_0$ meson mass. We have, for
example, the same  $\chi^2=29$ for the set of parameters corresponding to the
 $m_{f_0}=950$ MeV and  $m_{f_0}=985$ MeV. Note that for the single
$f_0$ meson the data on $K^-p\to\pi^+\pi^-(\Lambda\Sigma)$ reaction
are not sensitive as well.

 So, the spectra of $\pi\pi$ mesons obtained in the $\phi$ decay
$\phi\to\gamma\pi^0\pi^0$ is in good agreement with the other
experiments and the branching ratio of this decay support the hypothesis
of the four-quark nature of the $f_0$ meson.

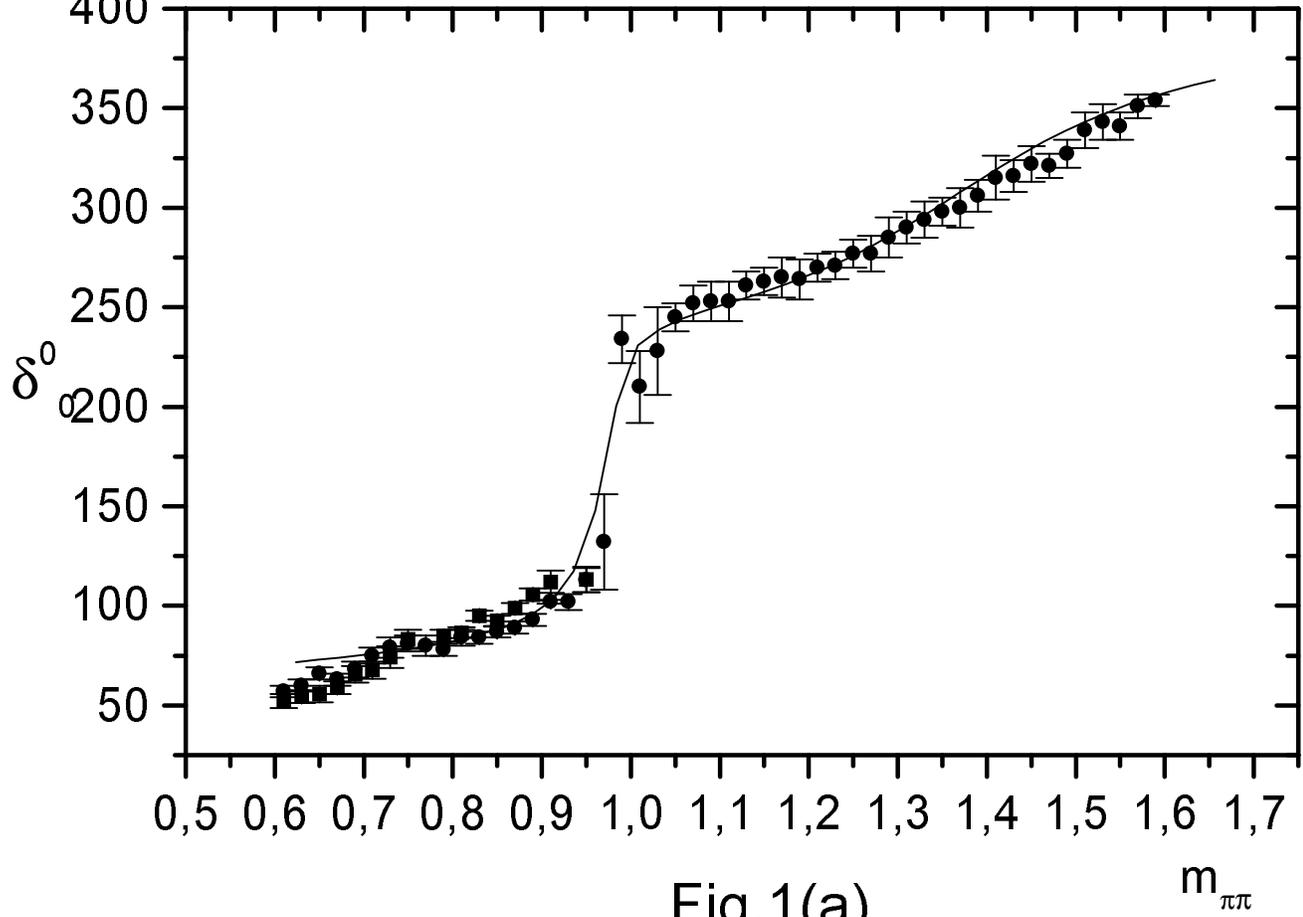
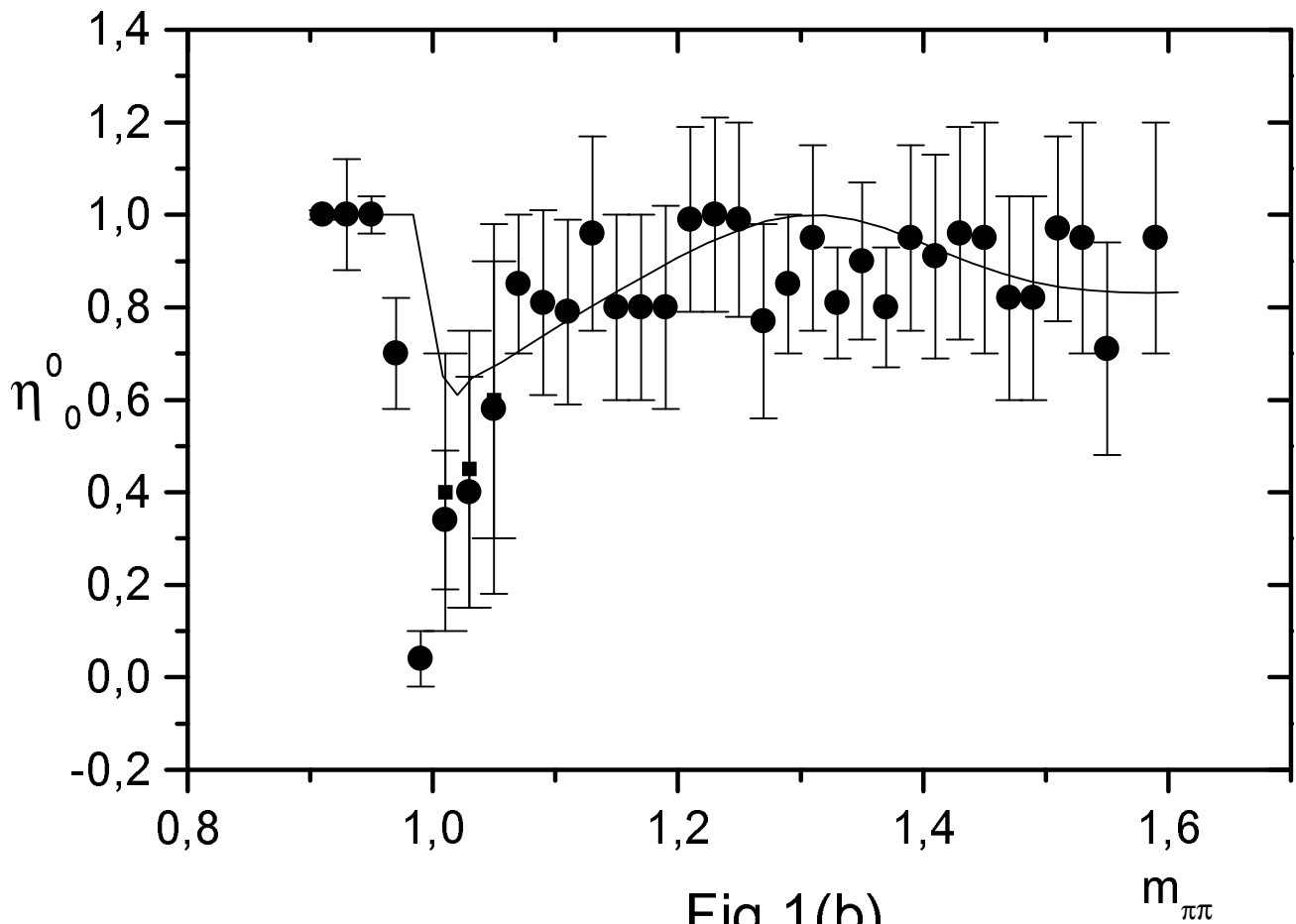
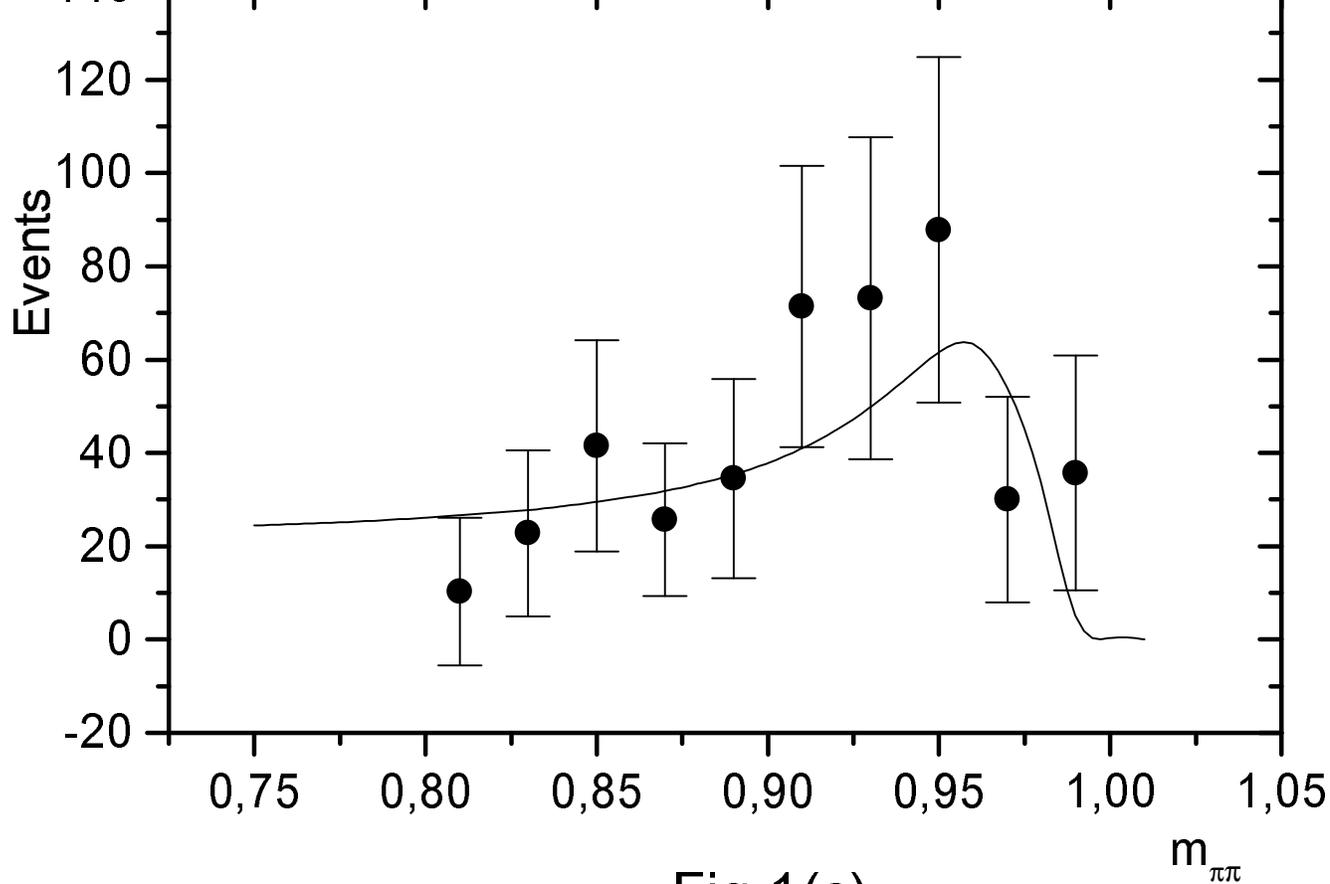
\begin{figure}
\caption{ The result of fitting. Parameters are $m_{f_0}=0.950$ GeV,
$m_{\sigma}=1.38$ GeV, $g^2_{f_0K^+K^-}/4\pi=2.25\ GeV^2$, 
$g^2_{\sigma\pi\pi}/4\pi=1.8\ GeV^2$,  $'=-0.34$, $R=3.5$ and
$\theta=43^{\circ}$. (a) The phase $\delta^0_0$. (b) The inelasticity
$\eta^0_0$. (c) The mass spectra of the reaction $e^+e^-\to\gamma\pi\pi$. }
\end{figure}

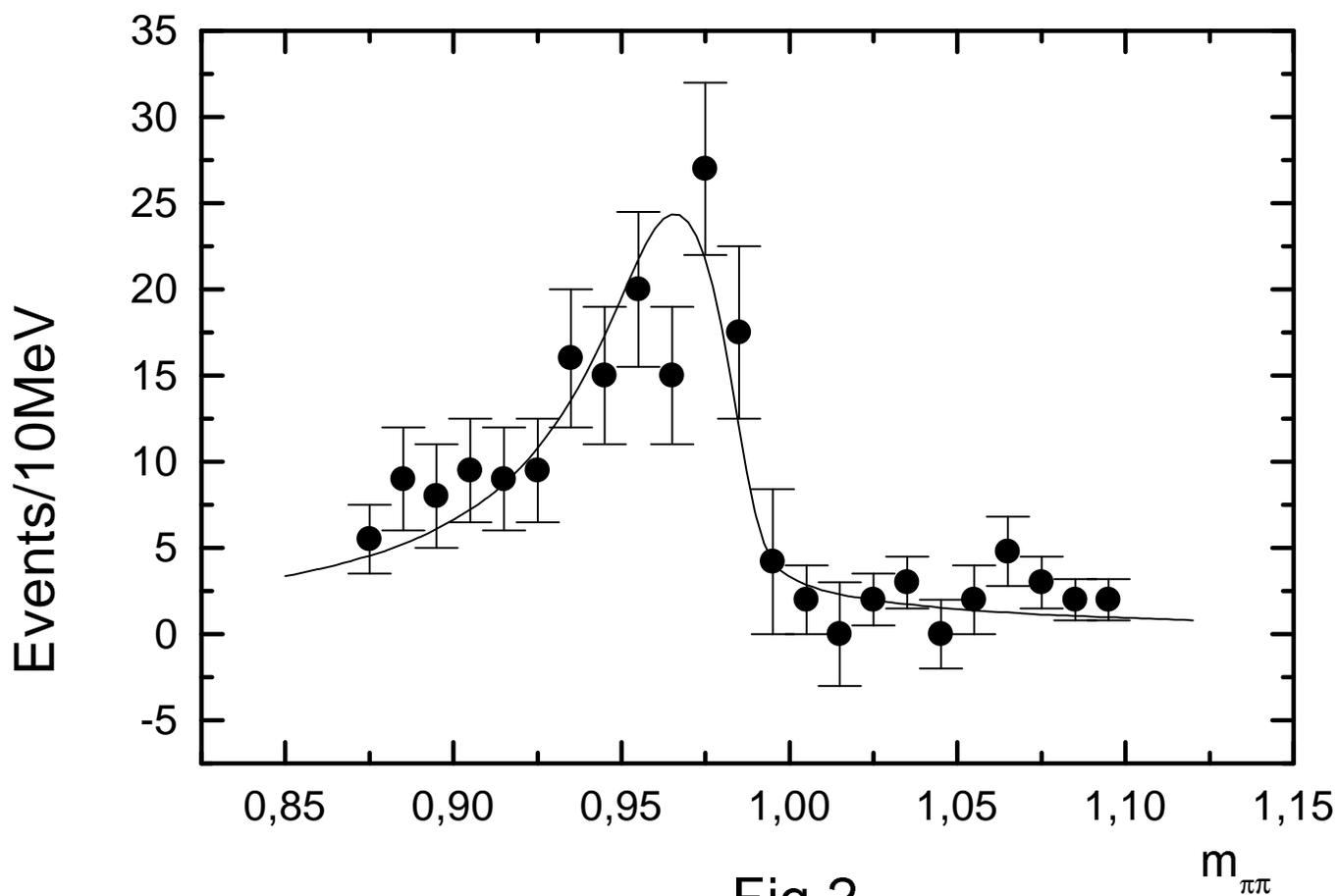
\begin{figure}
\caption{The mass spectra of the reaction $J/\psi\to\phi\pi^+\pi^-$.
C=9.0, $\xi=0.1$, the other parameters are the same as in Fig.1. }
\end{figure}

\begin{figure}
\caption{The mass spectra of the reaction  
$K^-p\to\pi^+\pi^-(\Lambda,\Sigma)$.
C=6.0, $\xi=0.1$, the other parameters are the same as in Fig.1. }
\end{figure}

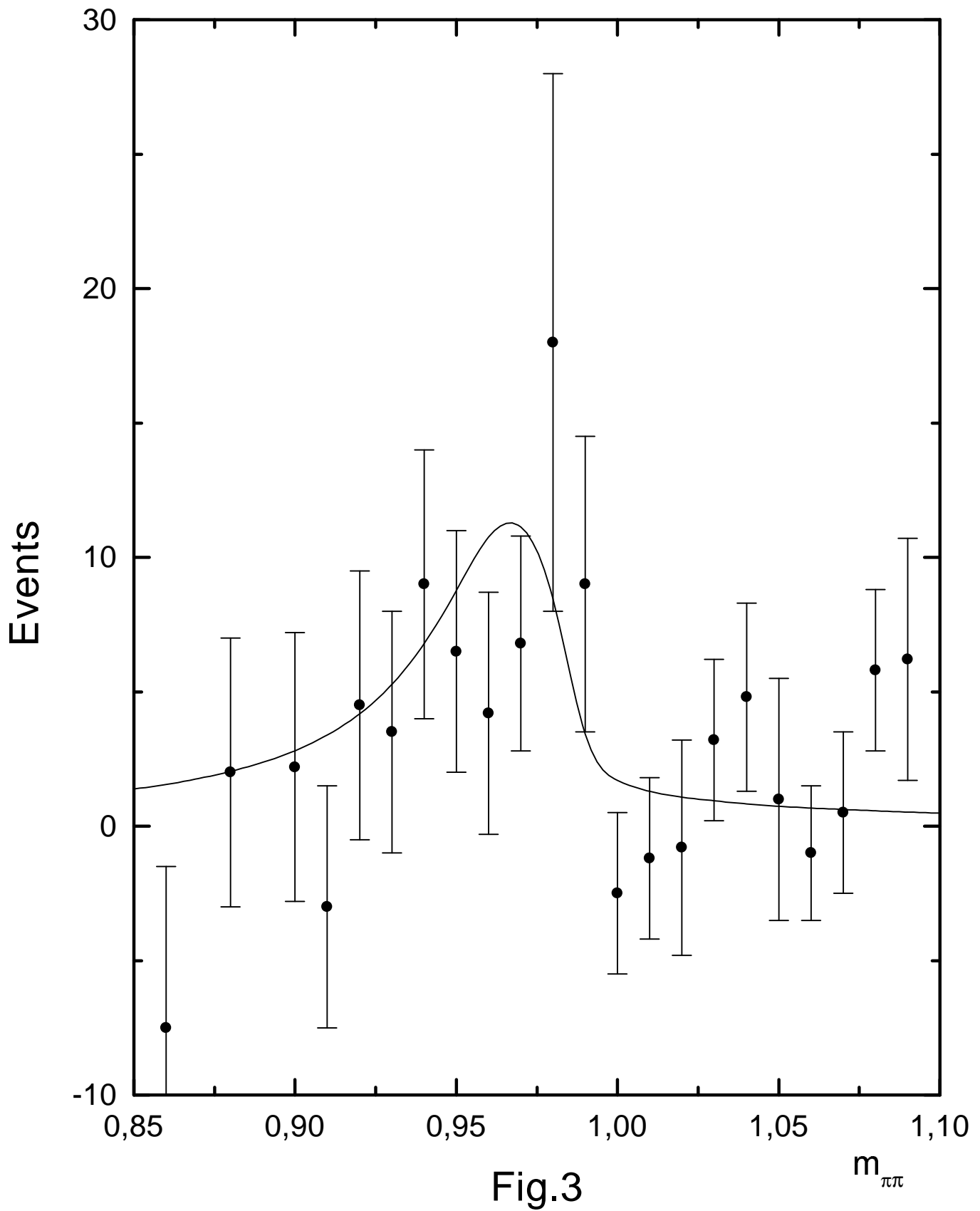
\end{document}